\documentclass[aps,prb,twocolumn,groupedaddress]{revtex4-1}
\usepackage{graphicx} 
\usepackage[dvipdfm]{hyperref}
\hypersetup{colorlinks=true,linkcolor=blue,
  implicit=true,breaklinks=true,pagebackref=true,backref=true,
bookmarks=true,bookmarksnumbered=true,hyperfootnotes=true,debug=true,
  naturalnames=false,citecolor=blue,pdfview=FitH,pdfstartview=FitH,hyperindex=true}
\usepackage{amsmath}
\usepackage{amssymb}
\usepackage{ifsym}
\begin{document}

\title{Superconductivity in half-Heusler compound TbPdBi}

\author{H. Xiao$^{1*}$}
\author{T. Hu$^{2,3\S}$}
\author{W. Liu$^{1,4}$}
\author{Y. L. Zhu$^{5}$}
\author{P. G. Li$^{5}$}
\author{G. Mu$^{2,3}$}
\author{J. Su$^{6}$}
\author{K. Li$^{1}$}
\author{Z. Q. Mao$^{5\sharp}$}

\affiliation{$^{1}$ Center for High Pressure Science and Technology Advanced Research, Beijing, 100094, China}
\affiliation{$^{2}$ State Key Laboratory of Functional Materials for Informatics, Shanghai Institute of Microsystem and Information Technology, Chinese Academy of Sciences, 865 Changning Road, Shanghai 200050, China}
\affiliation{$^{3}$ CAS Center for Excellence in Superconducting Electronics (CENSE), Shanghai 200050, China}
\affiliation{$^{4}$ Department of Physics, Zhejiang SCI-TECH University, Hangzhou, 310018, China}
\affiliation{$^{5}$ Department of Physics and Engineering Physics, Tulane University, New Orleans, LA 70018, USA}
\affiliation{$^{6}$ College of Chemistry and Molecular Engineering, Peking University, Beijing 100871, China}

\date{\today}
\begin{abstract}
We have studied the half-Heusler compound TbPdBi through resistivity, magnetization, Hall effect and heat capacity measurements. A semimetal behavior is observed in its normal state transport properties, which is characterized by a large negative magnetoresistance below 100 K. Notably, we find the coexistence of superconductivity and antiferromagnetism in this compound. The superconducting transition appears at 1.7 K, while the antiferromagnetic phase transition takes place at 5.5 K. The upper critical field $H_{c2}$ shows an unusual linear temperature dependence, implying unconventional superconductivity. Moreover,  when the superconductivity is suppressed by magnetic field, its resistivity shows  plateau behavior, a signature often seen in topological insulators/semimetals. These findings establish TbPdBi as a platform for study of the interplay between superconductivity, magnetism and non-trivial band topology.
\end{abstract}


\maketitle
\subsection{Introduction}

The large family of ternary half-Heusler compounds with non-centrosymmetric structure, formulated as XYZ (X = rare earth elements, Y = transition-metal elements, Z = main-group elements), has recently attracted a great deal of interests.\cite{Al-Sawai2010, Xiao2010, Chadov2010, Lin2010} In particular, the $R$PdBi and $R$PtBi ($R$ = rare earth) half Heusler series have shown to be an interesting platform for the study of unconventional superconductivity. For instance, YPtBi and LuPtBi have been reported to be superconducting,\cite{Butch2011, Bay2012, Meinert2016, Kronenberg2016, Pavlosiuk2016, Tafti2013, Hou2015, Liu2011, Mun2016} (their $T_c$ values are 0.77 K and 1 K respectively) even though they have a surprisingly low carrier concentration, i.e. $n=$ 10$^{18}$-10$^{19}$ cm$^{-3}$.\cite{Butch2011, Bay2012, Tafti2013} There have been compelling evidences which show the superconductivity in these compounds are unconventional. The low-temperature penetration depth measurements on YPtBi has revealed that its superconducting gap has nodes.\cite{HyunsooKim2016}  In addition, the unusual linear temperature dependence of the upper critical field points to an odd parity component in the superconducting order parameter, in accordance with the predictions for non-centrosymmetric superconductors. \cite{Bay2012} Due to strong spin-orbital coupling, the superconducting state of YPtBi is believed to have a mixture of a conventional pairing state and high-angular momentum paring states.\cite{Brydon2016, Yang2017, Savary2017, Venderbos2018, Boettcher2018, Roy2018} For LuPtBi, a surface nodal superconducting state has been observed with its $T_c$ being much higher  than that in the bulk.\cite{A.Banerjee2015}

In this paper, we report resistivity, magnetization, Hall effect and heat capacity measurements on the half Heusler compound TbPdBi. For the first time, we observed superconductivity in this compound with a onset temperature of $T_c=$ 1.7 K, besides the antiferromagnetic transition at $T_N=$ 5.5 K. Unlike other half-Heusler superconductors which feature semi-metallic normal states with large positive magnetoresistance, the superconductivity of TbPdBi is connected with an unusual normal state characterized by a large isotropic negative magnetoresistance. Regardless of this difference, TbPdBi exhibits a linear temperature dependence in upper critical field $H_{c2}$, similar to other half-Heusler superconductors, suggesting TbPdBi also possesses unconventional superconductivity. When its superconductivity is suppressed by magnetic field, its resistivity as a function of temperature shows a plateau behavior, suggesting the possible presence of non-trivial band topology. These results establish TbPdBi as an intriguing platform for the study of the interplay between unconventional superconductivity, magnetism and non-trivial band topology.

\subsection{Experimental Details}
Single crystals of TbPdBi were grown using Bi flux. We have performed single-crystal X-ray diffraction (SXRD) measurements on TbPdBi. The data were collected at 293(2)K on a Rigaku XtaLAB PRO 007HF(Mo) diffractometer, with Mo K$\alpha$ radiation ($\lambda=$ 0.71073 $\AA$). Data reduction and empirical absorption correction were performed using the CrysAlisPro program. The structure was solved by a dual-space algorithm using SHELXT program. Final structure refinement was done using the SHELXL program by minimizing the sum of squared deviations of F$^{2}$ using a full-matrix technique. Table 1 summarizes the detailed structural parameters extracted from the structural refinement, which shows the sample used in our study indeed has a cubic $F\overline{4}3m$ crystal structure. The occupancy of each element obtained from the refinement is close to 1, suggesting the composition of our synthesized compound is close to the stoichiometric ratio, i.e. TbPdBi. For transport measurements, the sample was first sanded and then cut into small pieces. The thickness of the sample used
is about 35 $\mu$m. The resistivity is measured down to 50 mK by using a dilution refrigerator in a physical properties measurement system (PPMS). The dc susceptibility was measured down to 2 K. Heat capacity were measured by a relaxation time method.

\begin{table}[]
\centering
\caption{Structural parameters of TbPdBi determined by single crystal XRD measurements at 293(2) K. Space group: $F\overline{4}3m$  (No. 216). Lattice parameters: $a=$ 6.65310(10) $\AA$, $b=$ 6.65310(10) $\AA$, $c=$ 6.65310(10) $\AA$, $\alpha=\beta=\gamma¦Á=$ 90$^{o}$. $R_{1}=$ 0.0351; $wR_{2} =$ 0.0836; $U_{eq}$ is defined as one-third of the trace of the orthogonalized $U_{ij}$ tensor ($\AA^{2}$).}
\label{my-label}
\begin{tabular}{|l|l|l|l|l|l|l|}
\hline
Atom & Wyckoff. & Occupancy. & x   & y   & z   & U$_{eq}$   \\ \hline
Bi   & 4b       & 1          & 1/2 & 1/2 & 1/2 & 0.0089(11) \\ \hline
Tb   & 4a       & 1          & 0   & 0   & 0   & 0.0112(16) \\ \hline
Pd   & 4d       & 1          & 3/4 & 3/4 & 3/4 & 0.013(2)   \\ \hline
\end{tabular}
\end{table}

\subsection{Results and Discussion}
 Figure 1(a) shows the temperature $T$ dependent resistivity $\rho$ measured under different applied magnetic fields ($H=$ 0, 1, 3, 5, 9 T). As cooling down from room temperature, the resistivity demonstrates a semiconductor-like behavior above certain temperature $T_{peak}$. Below that, it shows a metallic behavior. This behavior is characteristic of semimetals or narrow-gap semiconductors as observed previously in half-Heusler compound.\cite{Gofryk2011, Nakajima2015} The position of $T_{peak}$, marked by a downward arrow, shifts to higher temperature with increasing magnetic field, which is summarized in inset to Fig. 1(a). At higher temperatures ($T>$ 100 K), the resistivity curves measured in different $H$ merge into one single curve, while large negative magnetoresistivity (MR) is observed at low temperatures ($T<$ 100 K). This can be seen more clearly from Fig. 1(b) and its inset, which plots the $T$ dependence of $\rho$(9T)/$\rho$(0T)-1 and the $H$ dependence of $\rho(H)$/$\rho$(0T)-1, respectively.

 The large negative MR (with a magnitude up to 80$\%$) is a remarkable signature, contrasted with the nearly zero MR above $T_{peak}$. However, it is not clear yet about the origin of the negative MR and further study is needed to understand it. Note that for ordinary non-magnetic metal, the MR is usually weak and positive. In half-Heusler compounds, the MR is found to be positive and large. For example, in LuPtBi, positive MR as large as 3200$\%$ is reported.\cite{Hou2015} Negative and high anisotropic MR is reported in Weyl semimetals, such as TaAs-class materials, and has been regarded as the most prominent transport signature caused by the chiral anomaly effect.\cite{Huang2015} However, our observation of the negative MR in TbPdBi  is nearly independent of field orientation. Thus the negative MR observed in present case can not be understood in terms of any existing model.

\begin{figure}
\centering
\includegraphics[trim=0cm 0cm 0cm 0cm, clip=true, width=0.45\textwidth]{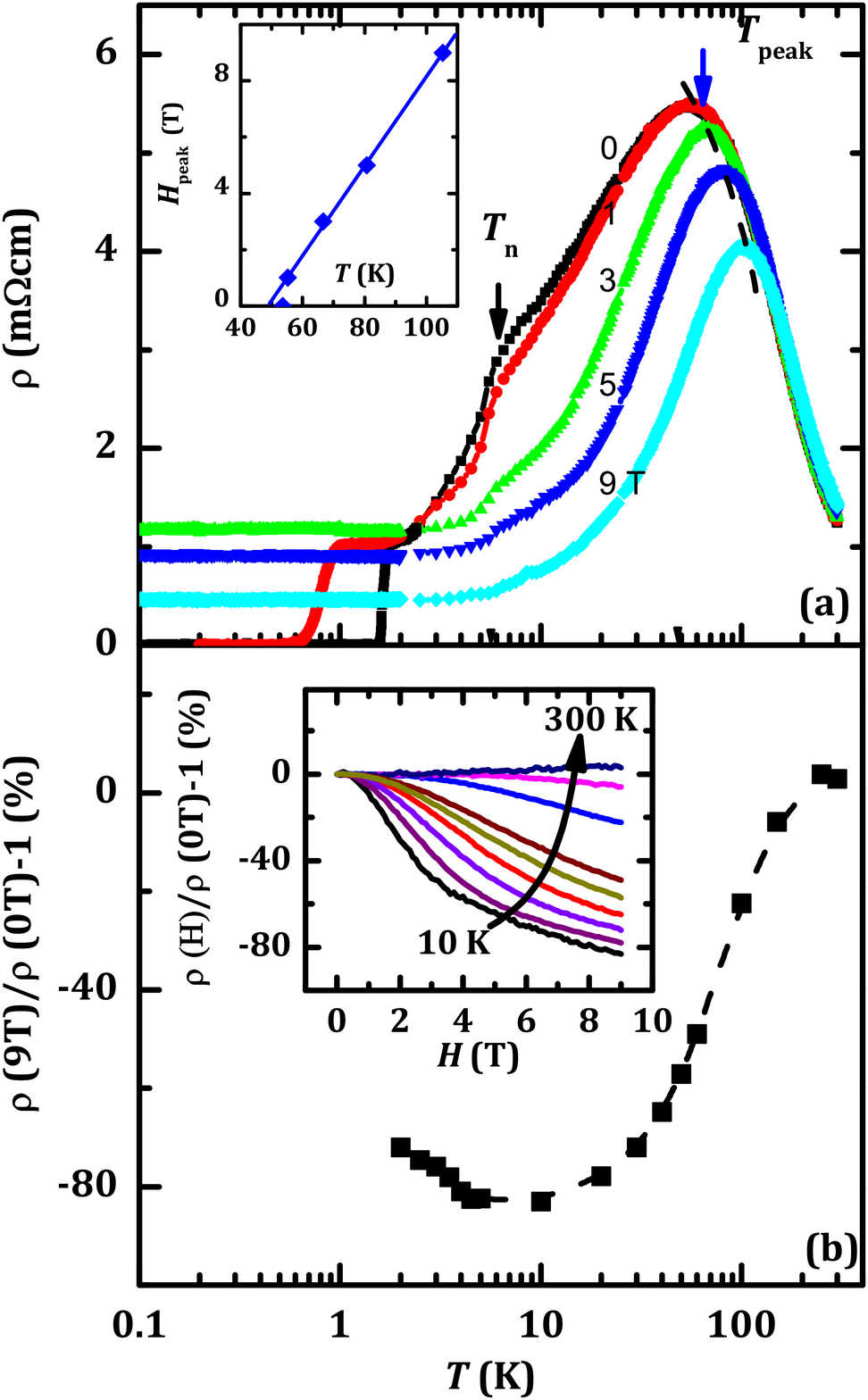}
\caption{\label{fig:magnetoresistance} (a) The resistivity $\rho$ vs. temperature $T$ data for TbPdBi from 50 mK to 300 K under applied magnetic field $H=$ 0, 1, 3, 5, 9 T.  Inset: the $T$ dependence of the resistivity peak in different magnetic field, $H_{peak}$. (b) The magnetoresistivity $\rho(9T)/\rho(0T)-1$ vs. temperature $T$. Inset shows $\rho(H)/\rho(0T)$-1 vs. $H$ at different temperatures, $T=$ 2, 10, 20, 30, 40, 50, 60, 100, 150, 300 K.  }
\end{figure}

\begin{figure}
\centering
\includegraphics[trim=0cm 0cm 0cm 0cm, clip=true, width=0.45\textwidth]{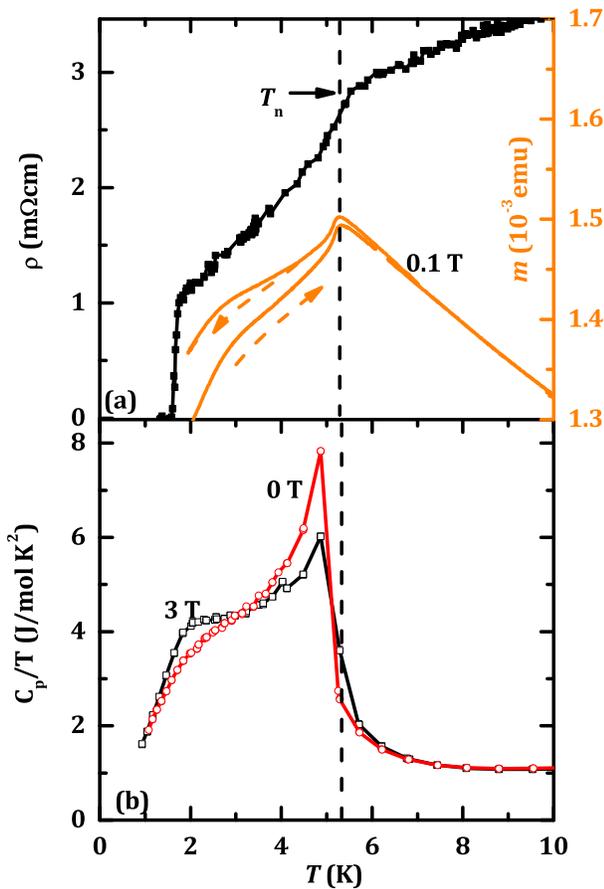}
\caption{\label{fig:magnetoresistance} (a) Left axis: The low temperature part of the $\rho$  vs. $T$ curve at zero magnetic field. Right axis: Magnetization measurements on TbPdBi with applied magnetic field $H=$ 1 kOe in zero filed cool (ZFC) and field cool (FC) conditions. (b) The temperature $T$ dependence of the specific heat ratio $C_P/T$ at $H=0$ and $H=3$ T. }
\end{figure}

Below $T_{peak}$, the resistivity curve shows a kink at 5.5 K,  which can be seen more clearly from the enlarged part of the low temperature resistivity curve (Fig. 2(a), left axis). Such a resistivity kink is due to an antiferromagnetic (AFM) phase transition previously determined by neutron diffraction measurements.\cite{Nakajima2015} The magnetization $M$ vs. $T$ curves measured at $H =$ 1 kOe in both zero field cooled (ZFC) and field cooled (FC) conditions are also shown in Fig. 2(a) (right axis), which suggest an AFM transition at $T_N=$ 5.5 K. Below $T_N$, the magnetization shows irreversibility, which may be caused by moment canting. Note that below $T_N$, the magnitude of the magnetoresistivity $\rho$(9T)/$\rho$(0T)-1 decreases with decreasing temperature, although it remains negative (see Fig. 1(b)).

With further decreasing temperature, the resistivity drops sharply at 1.7 K, down to zero at 1.58 K, signaling an onset superconducting transition at 1.7 K. The $T_c$ of 1.7 K is almost the same as that of LuPdBi which was reported to have the highest superconducting transition temperature among the superconductors found in the half Heusler family or other noncentrosymmetric systems.\cite{Xu2014} Although TbPdBi was previously studied, its superconductivity was not reported.\cite{Nakajima2015}  Previous transport measurements showed its resistivity exhibits a tendency of drop at about 0.5 K, but does not decreases to zero.\cite{Nakajima2015} This implies the sample used in our study somewhat differs from the sample used in previous work. In order to clarify such a possible sample dependence of superconductivity, we have examined several samples from different batches and found all of them show superconductivity. We also compared the transport measurements on the samples whose leads are prepared using  silver paste and silver epoxy respectively. The silver paste did not require baking, while the silver epoxy did. Both samples also showed the same superconductivity, which excludes the possibility that the superconducting phase we observed in TbPdBi is  induced by heating. One possible reason for the difference between our  sample and the reported one \cite{Nakajima2015} is that the reported sample likely involves non-stoichiometry, causing inhomogeneous superconductivity. The tendency of resistivity drop below 0.5 K observed in the reported sample is indeed a signature of inhomogeneous superconductivity. Note that recent penetration depth measurements also verified the superconductivity of TbPdBi.\cite{Radmanesh2018}

\begin{figure}
\centering
\includegraphics[trim=0cm 0cm 0cm 0cm, clip=true, width=0.45\textwidth]{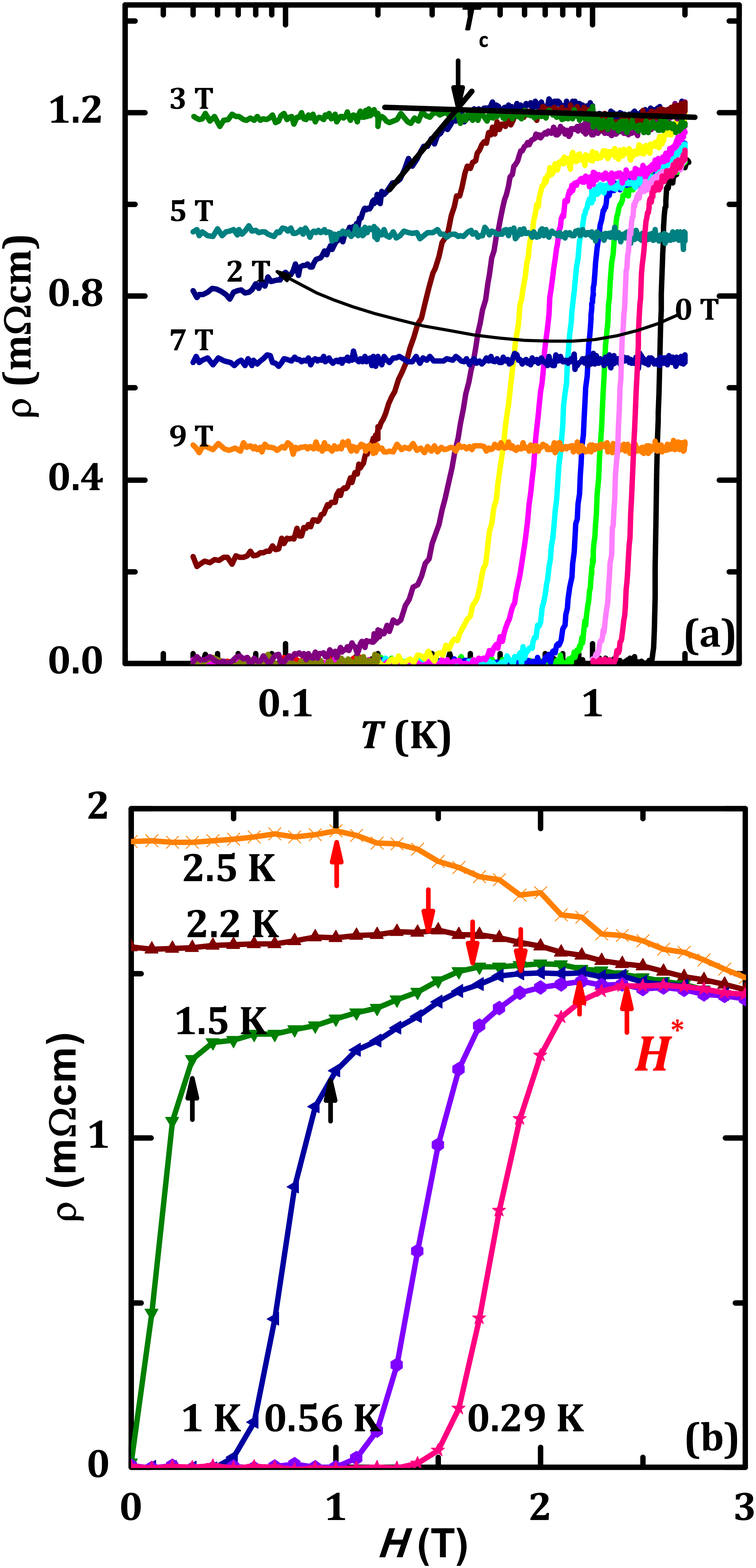}
\caption{\label{fig:magnetoresistance} (a) The resistivity $\rho$ vs. temperature $T$ for TbPdBi measured in a dilution refrigerator with applied magnetic field $H=$ 0, 0.2, 0.4, 0.6, 0.8, 1, 1.2, 1.4, 1.6, 1.8, 2, 3, 5, 7, and 9 T.
(b) The resistivity $\rho$ vs. magnetic field $H$ for TbPdBi at different temperatures, $T=$ 0.29, 0.56, 1, 1.5, 2.2, and 2.5 K.}
\end{figure}

 We also performed specific heat measurements on the TbPdBi sample. Figure 2(b) shows the temperature dependence of the specific heat ratio, $C_P/T$, measured at $H=0$ and $H=3$ T. From the zero field specific heat data, it is found that there is a sharp jump at $T_N=$ 4.86 K, which is coincident with the antiferromagnetic phase transition probed by resistivity and magnetization measurements. The magnitude of the jump is in the order of J/mol K$^{2}$,  consistent with the previous report, \cite{Nakajima2015} suggesting a huge release of magnetic entropy. With $H=3$ T, the peak position of $C_P/T$ remains unchanged but the magnitude of the peak gets suppressed. In addition, the 0 T data shows a humplike anomaly at lower temperatures, which is likely to originate from the change of spin structure. However, we did not observe a clear superconducting anomaly in $C/T$ at $T_c$, similar to the scenario seen in other half Heusler superconductors such as YPtBi\cite{Pavlosiuk2016} and HoPtBi.\cite{Pavlosiuk2016a} This can possibly be attributed to small effective mass of quasi-particles, thus resulting in electronic specific heat anomaly being too small to be observed.

Figure 3(a) shows the $\rho$ vs. $T$ curves measured under different applied magnetic field $H=$ 0, 0.2, 0.4, 0.6, 0.8, 1, 1.2, 1.4, 1.6, 1.8, 2, 3, 5, 7, and 9 T below 2 K. With increasing magnetic fields, the superconducting transition temperature is gradually suppressed to zero and the transition width becomes broader. The onset of the superconducting transition temperature $T_c^{onset}$ is defined as the cross point of the two extrapolated straight lines, as shown in Fig. 3(a). In zero magnetic field, $T_c^{onset}$ is determined to be 1.7 K. Based on these data, we obtain the temperature dependence of the upper critical field $H_{c2}$, as shown in Fig. 4(a) (circles). Note that $H_{c2}$  shows almost linear behavior in the whole measured temperature range and there is no sign of saturation at low temperatures, similar to what is observed in YPtBi.\cite{Butch2011}

The value of $H_{c2}$ at 0 K estimated from linear extrapolation is 2.4 T. Here we can estimate the superconducting coherence length at zero temperature, $\xi=(\frac{\Phi_0}{2\pi H_{c2}})^{1/2}=$ 12 nm. Note that the value of $H_{c2}$ for TbPdBi is comparable with that of other $R$PdBi/$R$PtBi superconductors. For example, $H_{c2}(0)$ is 2.2 T for LuPdBi \cite{Xu2014} and 1.5 T for YPtBi.\cite{Butch2011}
We also evaluate the orbital limiting field using the weak-coupling Werthamer-Helfand-Hohenberg (WHH) formula in the clean limit, $H_{orb}=0.69T_c[-dH_{c2}/dT]_{T_c}=$ 1.8 T. The Pauli limiting filed $H_p=\Delta/(\sqrt{2}\mu_B)$ where $\Delta=1.76 k_BT_c$ can be estimated to be 3.2 T. Since $H_{orb}<H_{c2}<H_{P}$, superconductivity in TbPdBi is orbital limited. But the fact that $H_{c2}$ is larger than the weak-coupling WHH estimation of $H_{orb}$ indicates that spin-orbital coupling is important in this material. In addition,
the linear temperature dependence of $H_{c2}$ suggests an unusual superconducting state. In the absence of inversion center, this may point to a possible mixed singlet-triplet pairing state.\cite{Tafti2013}

\begin{figure}
\centering
\includegraphics[trim=0cm 0cm 0cm 0cm, clip=true, width=0.5\textwidth]{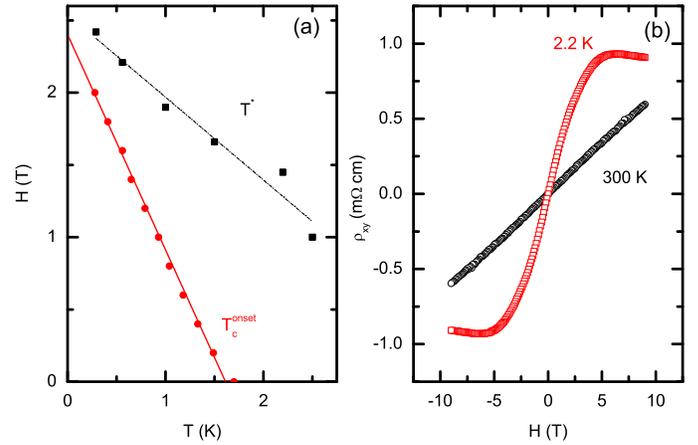}
\caption{\label{fig:magnetoresistance} (a) The magnetic field $H$ vs. temperature $T$ phase diagram. The circles represent the onset superconducting transition temperature $T^{onset}_c$. The squares denotes $T^{*}$, the crossover temperature of the positive MR to negative MR behavior at low temperatures. (b) The hall resistivity $\rho_{xy}$ vs. magnetic field $H$ at $T=$ 2.2 and 300 K.}
\end{figure}

It is interesting to note that a resistivity plateau emerges at low temperatures when the superconductivity is completely suppressed above $H=3$ T (see Fig. 3(a)). For a topological insulator (TI), the surface  which is in contact with air is metallic whereas the bulk is insulating, as a result of time reversal symmetry protecting the metallic surface modes of topological insulators. The transport signature of such a surface state is a plateau that arrests the exponential divergence of the insulating bulk with decreasing temperature. A resistivity plateau is reported in Bi$_2$Te$_2$Se, \cite{Ren2010} SmB$_6$,\cite{Kim2013} LaSb,\cite{Tafti2016}, TaSb$_2$,\cite{Li2016} and also in similar half Heusler compound LuPtBi.\cite{Hou2015} Hence, the resistivity plateau observed in TbPdBi implies that its electronic band structure involves non-trivial band topology. Further band structure calculations and ARPES measurements are needed to reveal its nature.

Fig. 3(b) shows the $H$ dependence of the $\rho$ at several selected temperatures, $T=$ 0.29, 0.56, 1, 1.5, 2.2 and 2.5 K. Note that there is a crossover from positive MR to negative MR behavior at $H^{*}(T^{*})$, which disappears at higher temperatures. Fig. 4(a) (squares) shows the magnetic field dependence of $T^{*}$, which increases with decreasing magnetic field. The origin of $H^{*}(T^{*})$ (position of MR peak) and its relationship to the superconductivity is not clear yet which requires further study.

The Hall resistivity $\rho_{xy}$ vs magnetic field $H$ at $T=$ 2.2 and 300 K is plotted in Fig. 4(b). At $T=$ 300 K, the linear dependence of $\rho_{xy}$ on the magnetic field indicate that one type of charge carrier dominates the transport properties at this particular temperature. Based on the one-carrier model, the carrier density $n$ is then estimated to be 9.43$\times$10$^{18}$cm$^{-3}$, comparable with other half-Heusler compounds.\cite{Xu2014, Butch2011, Nakajima2015} Such a low carrier density might explain why the specific heat data do not exhibit a discernible signature of $T_c$. At low temperatures, $T=$ 2.2 K, $\rho_{xy}$ is no longer linearly dependent on $H$, suggesting more complicated band structure. This is different from LuPdBi, where $\rho_{xy}$ is linear in $H$ at both $T=$ 2 K and $T=$ 300 K.\cite{Xu2014}

\subsection{Summary}

In summary, we report superconductivity with $T_c$ of 1.7 K in antiferromagnetic half-Heusler compound TbPdBi, which has an unusual normal state with large negative magnetoresistivity. The resistivity plateau at low temperature under magnetic field suggests possible non-trivial band topology. The upper critical field $H_{c2}$ shows unusual linear dependence on temperature, implying unconventional superconductivity. Thus, TbPdBi provides a new platform to study the interplay of topological states, superconductivity and magnetism.

\subsection{Acknowledgments}
 We thanks Peijie Sun from IOP, CAS, Haiyan Zheng, Xiaohuan Lin from HPSTAR, and Hui Zhang from SIMIT for helpful discussions. Work at HPSTAR was supported by NSAF, Grant No. U1530402. Work at SIMIT was supported by NSFC, Grant No. 11574338. The efforts of sample growth and a part of data analysis at Tulane were supported by the U. S. Department of Energy under EPSCOR Grant No. DE-SC0012432 with additional support from Louisiana Board of Regents. G.M. acknowledge the support from the Youth Innovation Promotion Association of the Chinese Academy of Sciences (NO. 2015187). K.L. acknowledge the support from NSFC 21771011.

$^{*}$ hong.xiao@hpstar.ac.cn  $^{\S}$ thu@mail.sim.ac.cn  $^{\sharp}$ zmao@tunlane.edu
%


\end{document}